\documentclass[cite1,clbiba,12pt,psfig]{article}
\usepackage{epsfig}
\usepackage{float}
\usepackage{graphicx}
\usepackage{overcite}
\usepackage{color}
\usepackage[table,xcdraw]{xcolor}
\usepackage{chapterbib}

\usepackage{fancyhdr}
\usepackage{hyperref}
\usepackage{authblk}

\title{Novel force field of {[Bmim][Nf$_2$T]} and its tranferability in a mixture with water.   }
\date{}
\author[1]{\small Ra\'ul Fuentes-Azcatl \\
  \href{mailto:rfuentes@ifuap.buap.mx}{rfuentes@ifuap.buap.mx}}
\author[2]{\small Minerva Gonz\'alez-Melchor \\
  \href{mailto:minerva@ifuap.buap.mx}{minerva@ifuap.buap.mx}}
\affil[1]
{\footnotesize Instituto de F\'isica ''Luis Rivera Terrazas'', Benem\'erita Universidad Aut\'onoma de Puebla, Apdo. Postal J-48, Puebla, 72570, M\'exico\\}

\newpage 

\begin{document}

\newpage

\maketitle
\begin{abstract}
In this work, a new force field is presented for the {ionic liquid} 1-butyl-3-methylimidazolium - bis(trifluoromethylsulfonyl)imide, [Bmim][Nf$_2$T]. As a part of the $\epsilon$ force field, the acronym IL/$\epsilon$ is used to refer to this ionic liquid. This new force field reproduces the dielectric constant, the density, and the entalphy of vaporization, with an error of less than 3\%, being posible to generate new force fields for ionic liquids, based on the flexibility of the molecule. In addition, a study of the [Bmim][Nf$_2$T]-H$_2$O mixture is performed from pure IL to pure water, passing through various concentrations.\\

\end{abstract}


\section{Introduction}


The study of ionic liquids (IL) has been extensively analyzed and reported  by experimental techniques with the expectation that they could be applied as green solvents in the engineering of chemical reactions \cite{BADGUJAR20152}. Computer simulation methods and theoretical approaches have been widely used for studies in liquid phase of these systems \cite{PANDEY200638,CanongiaLopes2004,Roosen2008} and mixed with polar and non-polar solvents, where a critical point of the liquid-liquid equilibrium at ambient temperature has been observed \cite{Schroer2016, Fuentes-Azcatl2021}.

In computational theoretical chemistry, there is a wide variety of force fields(FF): from those where molecules are described by all their atoms to those that consider a coarse grain scheme and define a characteristic group of various atoms into a single pseudoatom, which reduces the degrees of freedom and speed up the calculations \cite{MARTINEZJIMENEZ2021115488}.

[Bmim][Nf$_2$T] is important because it is known to be non-toxic, so it can be applied in metabolite recovery, environmental biotechnology, and biosynthesis of high-value chemicals \cite{QUIJANO,Munoz,Morrish,Malinowski}.
Toxicity studies of IL (acute toxicity assays) are usually based on bioluminescence using microorganisms such as Vibrio fischeri or Photobacterium phosphoreum \cite{Romero,Garcia}, so, an adequate study of the mixtures with water is important.
The [Bmim][Nf$_2$T] is also relevant for those applications where ILs are employed in the form of thin films supported on solid surfaces, such as in microelectromechanical or microelectronic devices \cite{Bovio2009}.

After analyzing various models of IL's, it can be concluded that the available FFs do not reproduce all the experimental data, although a large percentage reproduces the density \cite{MARTINEZJIMENEZ2021115488}, the values produced for other properties are very dispersed and show large differences with respect to the experimental data. In the work of Acevedo et al. \cite{Doherty2017-st}, the charges of a previously proposed FF \cite{Sambasivarao2009-pt} were reparameterized with a factor of 0.8. Although various properties were improved, they are still far from the experimental data; here we will present this comparison.

Theoretical-computational calculations, specifically those carried out with molecular simulation methods, are currently an important route to understand the role played by molecular interactions on thermodynamic properties and thus, to obtain reliable data on the dynamic and structural properties of non-polarizable liquids. The FF of polar and nonpolar liquids are typically developed by combining electronic structure and molecular simulation calculations to obtain molecular geometries, partial charges of atoms and parameters to reproduce liquid properties, but they also include potentials that give freedom to describe bonds, angles and dihedrals, a procedure that is not uniquely defined.

The remainder of the  paper goes as follows: In Section II,  
the models are introduced. Section III
shows the simulation details; in Section IV, 
 the results are analysed, and finally, conclusions are presented in Section V.

\section{ Models}

The intramolecular energy of [Bmim][Nf$_2$T] ionic liquid is evaluated as the sum of contributions due to harmonic bond stretching, angle bending terms and a cosine series for torsional energetics.  The non-bonded interactions are described by Coulomb and Lennard-Jones (LJ) potentials. So, the FF parameters are the force constants, k$_r$ and k$_\theta$, the equilibrium bond and angle values, r$_0$ and $\theta_0$, respectively, and the Fourier coefficients $V$'s for the torsional energetics function of the angle $\phi$, all contained in the following equations,

\begin{equation}
\label{Kbond}
\sum {_{bonds}}=\sum\limits_{i}k _{b,i}(r_i - r _{0,i})^2
\end{equation}
\begin{equation}
\label{kangle}
\sum {_{angles}}=\sum\limits_{i}k _{\theta,i}(\theta_i - \theta_{0,i})^2
\end{equation}

\begin{equation}
\label{deltaH1}
\sum {_{torsion}}=\sum\limits_{i} [\frac{1}{2} V_{1,i}(1+cos \phi)+\frac{1}{2} V_{2,i}(1-cos 2\phi)+\frac{1}{2} V_{3,i}(1+cos 3\phi)+\frac{1}{2} V_{4,i}(1-cos 4\phi)].
\end{equation}


For the intermolecular potential between two molecules, the LJ and Coulomb interactions of non-polarizable models are used,

\begin{equation}
\label{ff}
u(r) = 4\epsilon_{\alpha \beta} 
\left[\left(\frac {\sigma_{\alpha \beta}}{r}\right)^{12}-
\left (\frac{\sigma_{\alpha \beta}}{r}\right)^6\right] +
\frac{1}{4\pi\epsilon_0}\frac{q_{\alpha} q_{\beta}}{r}
\end{equation}

\noindent where $r$ is the distance between sites $\alpha$ and $\beta$, $q_\alpha$ is the electric charge of site $\alpha$, $\epsilon_0$ is the permittivity of vacuum,  $\epsilon_{\alpha \beta}$ is the LJ energy scale, and  $\sigma_{\alpha \beta}$ is the repulsive diameter for an $\alpha-\beta$ pair. The cross interactions between unlike atoms are obtained using the Lorentz-Berthelot mixing rules,

\begin{equation}
\label{lb1}
\sigma_{\alpha\beta} = \left(\frac{\sigma_{\alpha\alpha} +
	\sigma_{\beta\beta} }{2}\right);\hspace{1.0cm} \epsilon_{\alpha\beta} =
\left(\epsilon_{\alpha\alpha} \epsilon_{\beta\beta}\right)^{1/2}.
\end{equation}

The new force field Il/$\epsilon$ was obtained by changing the angular constants of the harmonic potential k$_{\Theta}$. This is because the intramolecular interaction of the atoms has a component due to a restrictive force in their angular movement, and increasing it causes the restrictive force to decrease and atoms to vibrate more. If k$_{\theta}$ decreases in order to have a restrictive movement, it is enough to improve the density at various temperatures and reproduce the ${\Delta}$H at room temperature, as will be reported in the results of this work. Although there is freedom in the angle between the atoms of the anions and cations, it was restricted by 20$\%$ with respect to that reported by Sambasivarao et al. \cite{Sambasivarao2009-pt}. This FF is all atom, and with the new parameters k$_{\theta}$, the calculated properties are in agreement with the aforementioned experimental data. The value of 20$\%$ was found after analyzing a large number of k$_{\theta}$ values. We also studied the harmonic potential of the bonds, but it was noted that there was not a better description of the experimental data when $k_r$ was modified; therefore, in this work only the constants k$_{\theta}$ were modified.

\section {Simulation details}

The computational calculations were carried out by solving the equations of motion of particles by means of molecular dynamics (MD) simulations on the liquid phase. The simulations were performed using the isothermal-isobaric (NpT) ensemble \cite{tuckerman2010statistical} to obtain the density, the dielectric constant, the self-diffusion coefficient and the heat of vaporization of the system; for this purpose, the GROMACS package was used \cite{Van_Der_Spoel2005-oo}. The molecules are randomly located in the simulation cell
divided in each axis into slices of similar proportions, which generates planes perpendicular to each axis, and at the intersections of
these planes, a molecule is placed according to its center
of mass. This procedure is done with home-made programs.
A total of 500 molecules (250 cations and 250 anions) were used in the simulations of the IL and 864 for the  water-IL solution. The equations of motion were solved using the  leapfrog algorithm with a time step of 1 fs. Periodic boundary conditions were
used in the directions, and the bond distances were kept
rigid with the LINCS procedure, \cite{Hess2008-zk} when necessary.
The cutoff distance was equal to 1.0 nm for both the real
part of the Coulomb potential and the LJ interactions; in
addition, analytical long-range corrections were applied for the second one. The PME method \cite{1995JChPh.103.8577E} was applied to evaluate the reciprocal contribution with a grid of 0.34 nm for the reciprocal vectors with a spline of order 4. The Nosé–Hoover thermostat and Parrinello–Rahman barostat were applied with coupling times of 0.6 and 1.0 ps, respectively. The average properties were obtained after production runs of 50 ns, and the time to equilibrate the initial system in the liquid phase was 1 ns.

The surface tension was obtained from a NVT simulation
in which a parallelepiped cell was used, with L$_x$ = L$_y$ = 13.8 nm
and L$_z$ = 3 L$_x$ to avoid finite effects \cite{orea2005oscillatory, minervaST}, containing 5324
molecules. Periodic boundary conditions were applied in
all directions. The molecules are initially allocated in a
liquid slab surrounded by vacuum with two symmetrical
interfaces. 
The average components of the pressure tensor were
obtained for 30 ns after an equilibration period of 5 ns. The
densities of the two phases were extracted from the statistical averages of the liquid and vapor limits of the density
profiles \cite{doi:10.1063/1.469505}. The corresponding surface tension $\gamma$ on the planar interface was calculated from its mechanical definition \cite{doi:10.1063/1.469505}, according to
\begin{equation}
\label{lb2}
\gamma=0.5L_z[<P_{zz}>-0.5(<P_{xx}><P_{yy}>)],
\end{equation} 
where L$_z$ is the length of the simulation cell in the longest direction and P$_{\alpha\alpha}$ ($\alpha = x,y,z$) are the diagonal components of the pressure tensor. The factor 0.5 outside the squared brackets takes into account the two symmetrical interfaces in the system.

\section {Results}

\subsection{Il/$\epsilon$}
The new force field was obtained from a previous proposal by Canongia et al. \cite{CanongiaLopes2006,CanongiaLopes2008} by changing the constant k$_\theta$ of the angular harmonic potential. Such function in the intramolecular part models the restrictive force in the angular movement of molecules. An increase in k$_\theta$ causes an increase in the force; the frequency of oscillation also increases, and therefore, molecules vibrate more. So, by decreasing k$_\theta$ in order to relax the restrictive force, it can be enough to produce changes in the properties, improving the density, the dielectric constant, and the enthalpy of vaporization $\Delta$H$_{vap}$ at different temperatures. Although there is freedom in choosing the angle between the atoms of the anions and cations,when it was restricted by 20\% with respect to that reported by Canongia et al., \cite{CanongiaLopes2006,CanongiaLopes2008} the agreement between the aforementioned properties and the experimental data improved. The value of 20\% was found after analyzing a large number of k$_\theta$-values. In addition, the harmonic potential of the bonds was studied too, however, it was noted that modifications to k$_r$ did not produce a better description of the properties. Therefore, it was decided to only modify the constant k$_\theta$. This method, identified as the CO$_2$/$\epsilon$\cite{co2e}, has been previously used to parameterize the CO$_2$ molecule, improving the properties already mentioned when contrasting them with the experimental data.
\subsubsection{Density, $\rho$}

The density of the ionic liquid [Bmim][Nf$_2$T] has been widely studied and reported by various research groups, with different approaches to molecules, from those that describe all the atoms to those that, in a pseudo-atom, include the contribution of a set of atoms, mainly hydrogens, which are taken into account in a united atom\cite{MARTINEZJIMENEZ2021115488}.

\begin{figure}[H]
		\centering
	 {\includegraphics[clip,width=12cm,angle=-90]{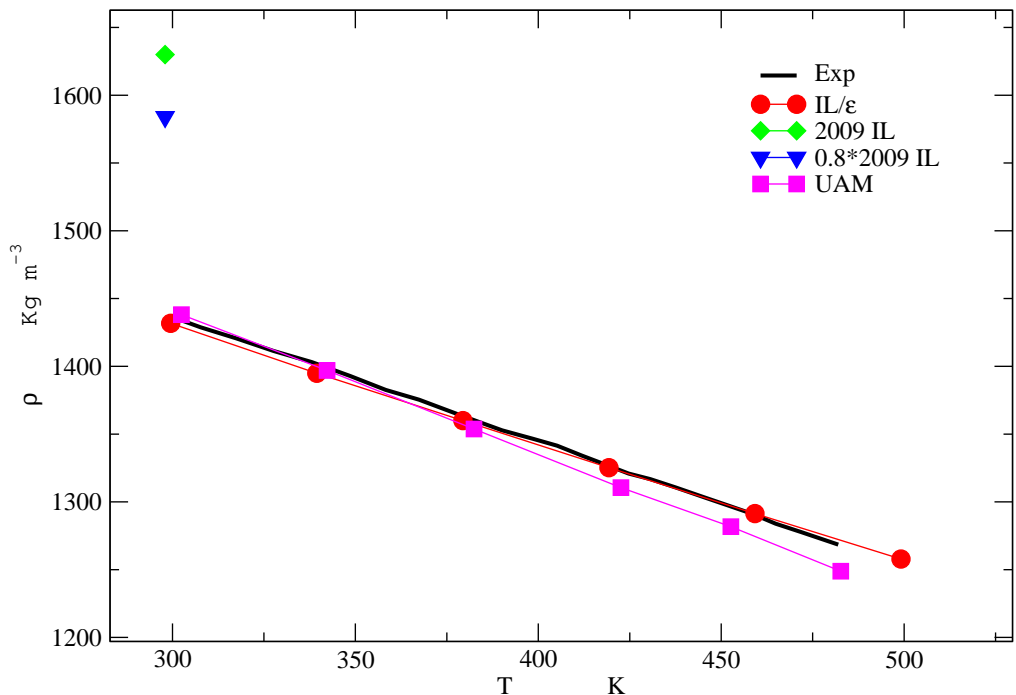}}
		
	 \caption{Density of the system [Bmim][Nf$_2$T] at different temperatures. Black triangles are the experimental data \cite{hamidova}. The red circles are the results of IL/$\epsilon$, the blue triangle is reported by Doherty et al. \cite{Doherty2017-st}, the green diamond is reported by Sambasivarao et al. \cite{Sambasivarao2009-pt}, and the magenta square by Mart\'inez-Jim\'enez et al. \cite{MARTINEZJIMENEZ2021115488}}
		\label{FigDens}
	\end{figure}

In this paper, we will compare our new IL/$\epsilon$ model with the all-atom models of Doherty et al. \cite{Doherty2017-st} and Sambasivarao et al. \cite{Sambasivarao2009-pt},  and with the newly parameterized united atom model developed by Mart\'inez-Jim\'enez et al. \cite{MARTINEZJIMENEZ2021115488}. The calculated density is shown in figure \ref{FigDens}, where an excellent agreement with experimental data is observed for all temperatures.

	
\subsubsection{Heat of vaporization, ${\Delta}H_{vap}$}

 An important characteristic of  ionic liquids is the vaporization enthalpy, its magnitude is generally greater than that of other molecular liquids due to the strong electrostatic interactions between the ions. Reproducing heats of vaporization, $\Delta$H$_{vap}$, is particularly important as it serves as an indicator of the average intermolecular interactions.

The calculation of $\Delta$H$_{vap}$ is obtained by means of the energetic difference between the liquid phase and the gas phase,

\begin{equation}
\label{deltaH2}
\Delta H_{vap} = \Delta H_{gas} - \textcolor{blue}{\Delta H_{liq},}
\end{equation}
which can be written as:
\begin{equation}
\label{deltaH3}
\Delta H_{vap} = E_{total}(gas)-E_{total}(liq) + RT.
\end{equation}

Experimental data has shown that ionic liquids go into
the vapor phase as an ion pair, \cite{doi:10.1021/jz301608c, Malberg2015} then the energy E$_{total} (gas)$ is calculated using a single ion pair and E$_{total} (liquid)$ is the
corresponding value in the condensed phase. In the simulation we have obtained the PV term and it is used in place of RT term.

\begin{table}
		\caption{Calculated and Experimental Heat of Vaporization
(kJ/mol) at 298 K.}

		\scalebox{1}[0.9]{
			\begin{tabular}{|c|c|c|c|c|c|c|c|c|}
				\hline		
				Exp.	&	2009IL	&	\% error	&	0.8*2009 Il	&	\% error	&	UAM	&	\% error	&	IL/$\epsilon$	&\% error	\\
				\hline	
				\hline	
				
134.7	&	190.4	&	41.3	&	154.8	&	14.9	&	50.68	&	62.4	&	134.1	& 0.4	\\

				\hline		
		\end{tabular}}
		\label{tableIR}
	\end{table}
The parameterization carried out by Doherty et al\cite{Doherty2017-st}, by reducing the general charge of the ionic liquid molecules reduced the error by 26.4\%, but even so the error is large. With the new force field the error was reduced to 0.4\% , which basically equals the calculated $\Delta H{_{vap}}$ to its experimental value, as shown in Table \ref{tableIR}.


\subsubsection{Surface tension, $\gamma$}
In the study carried out by  Doherty et al \cite{Doherty2017-st}, the reproduction of the surface tension improved: the value they reported has an error of 29.23\%. In this work that data is compared with the new force field IL/$\epsilon$ value.\\
The calculation of the surface tension in ionic liquids is fundamental due to its nature, since essentially, the tangential components of the pressure define it. In general, the surface tension is high within a wide range of temperatures. Another property related to this phenomenon is the vapor pressure, which is very low.

\begin{figure}[H]
		\centering
	 {\includegraphics[clip,width=12cm,angle=-90]{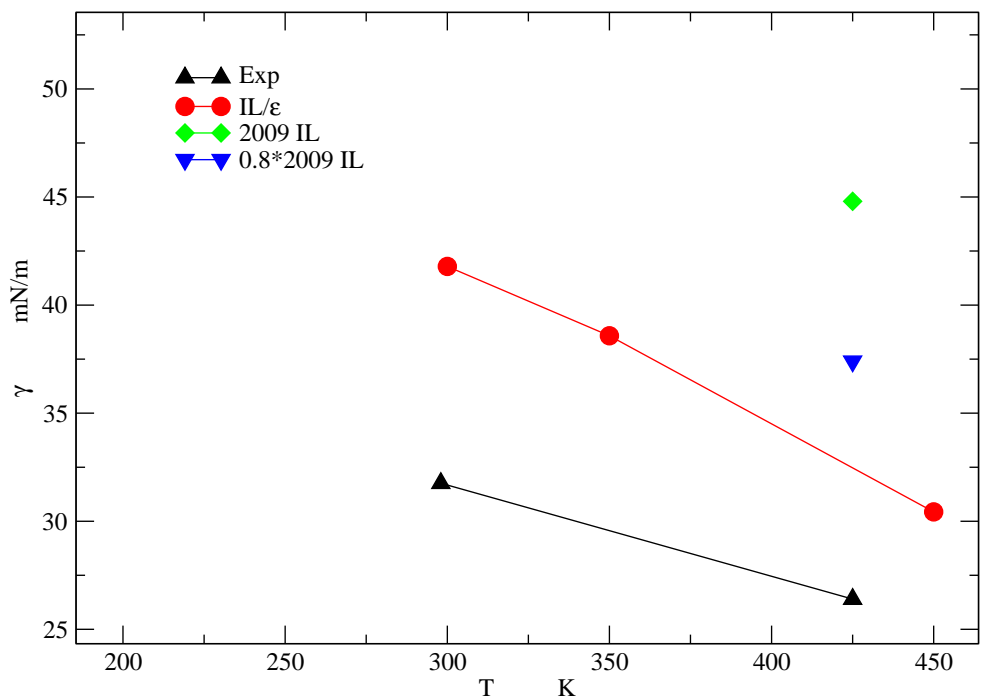}}
		
	 \caption{The surface tension of the system [bmim] [Tf2
N] at different temperatures. Black triangles are the experimental data \cite{hamidova}. the red circles are the result of IL/$\epsilon$, the blue triangle is reported by Doherty et al \cite{Doherty2017-st} , the green is reported by Sambasivarao et al \cite{Sambasivarao2009-pt} }
		\label{FigST}
	\end{figure}

	In the figure \ref{FigST}, the results of the new force field IL/$\epsilon$ are shown. The surface tension decreases as the temperature increases, the IL/$\epsilon$ force field reproduces the experimental trend but moves away from those data in a proportion of 30 \% at 298 K and 19.6 \% at 425 K while the vapor pressure remains low, as data in table \ref{tablePvap} shows, indicating strong cohesion between the ions and, as consequence, a low volatility.
	
	\newpage
\begin{table}
		\caption{Absolute vapor pressure at different temperatures of the IL/$\epsilon$.}

		\scalebox{1}[0.9]{
			\begin{tabular}{|c|c|c|}
				\hline
				
					Temp. 	&	$\mid Pvap\mid _{DM}$&	$Pvap_{Exp}$	\\
					K & bar& bar\\
				\hline
					\hline
300	&	0.146	&	-\\
350	&	0.142	&	-\\
450	&	0.094 &	7.1e-8 \\

				\hline		
		\end{tabular}}
\label{tablePvap}
	\end{table}
\newpage

\subsubsection{Dielectric constant, $\epsilon$.}
Ionic liquids are charged molecules and inherent to their study, an analysis of electrical behavior must be considered. A property governed by the interactions produced by the charges and their positions in space is the dielectric constant, which is calculated through the fluctuations in the total dipole moment of the system \cite{neumann1983dipole},

\begin{equation}
\label{diel}
\epsilon=1 + \frac{4}{3k_B TV}(<M^2>-<M>^2),
\end{equation}
where $k_B$ is the Boltzmann constant and T is the absolute temperature. The dielectric constant was obtained from simulations of 55 ns using the isotropic NPT ensemble, which are long compared to simulation times required to calculate other properties, such as the surface tension. The proper evaluation of the dielectric constant needs long simulations to ensure an average dipole moment of the system around zero\cite{Fuentes-Azcatl2014-ez} and the convergence of the property with time.

\begin{figure}[H]
\vspace{2.0 cm}
		\centering
	 {\includegraphics[clip,width=13cm,angle=0]{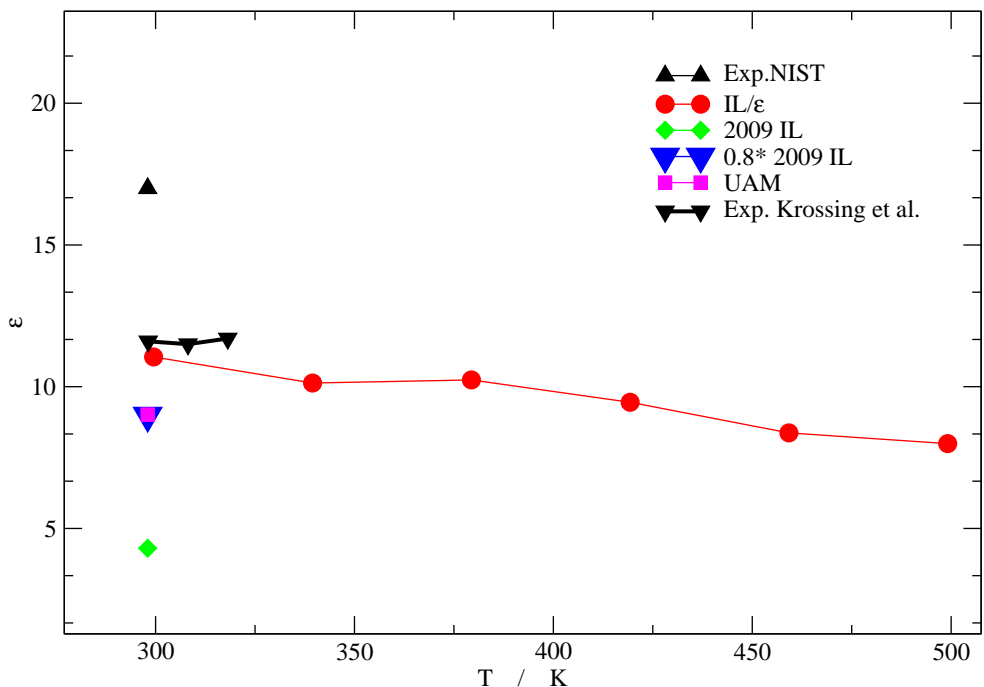}}
		
	 \caption{Dielectric constant of the system [Bmim] [Tf2N] at different temperatures. Black triangle up is the experimental data taken from NIST database \cite{NIST}, black triangle down is the experimental data reported by Krossing et al \cite{ Krossing2006-bc}. the red circles are the result of IL/$\epsilon$, the blue triangle  by Doherty et al \cite{Doherty2017-st}, the green diamon by Sambasivarao et al \cite{Sambasivarao2009-pt} , magenta square by Martínez-Jiménez et al \cite{MARTINEZJIMENEZ2021115488}, all were calculated in this work}
		\label{FigDiel}
		
	\end{figure}

The results of the dielectric constant, obtained from the molecular dynamic calculations, are shown in the figure \ref{FigDiel}. All the force fields studied here were compared by evaluating $\epsilon$ at 300 K and 1 bar. There is an important difference between the values reported at the NIST database \cite{NIST} and those reported in the work of Krossing et al\cite{ Krossing2006-bc}; taking into account the values of Krossing and coworkers, the 2009IL is 63.2\% far from the experimental data, the 0.8*2009IL is 22.4\% far equal to the UAM FF while the result produced by the IL/$\epsilon$ FF has a difference of 5\%. Although the dielectric constant is low in all FFs, the value here reported and calculated using the new Il/$\epsilon$ model is the closest to the experimental data. From the references found and analyzed, we noted that the dielectric constant is not reported. Since $\epsilon$ is a very important property due to the nature of the Coulombic interactions in these fluids, our work contributes filling this gap in the description of IL's.

\begin{figure}[H]
\vspace{2.0 cm}
		\centering
	 {\includegraphics[clip,width=12cm,angle=-90]{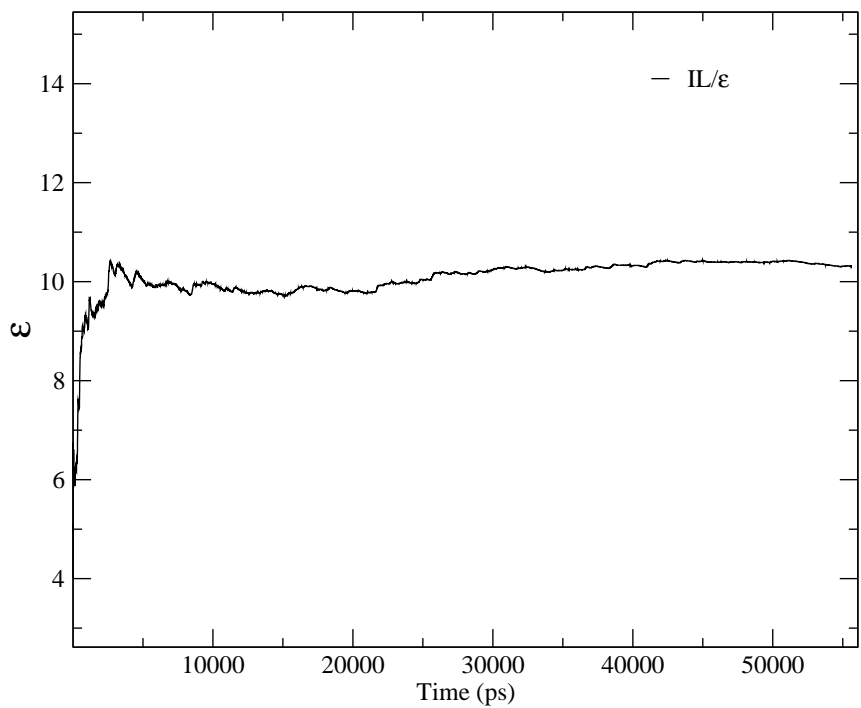}}
		
	 \caption{Dielectric constant  of IL/$\epsilon$. evaluation of the dielectric constant after 55 nanoseconds.}
		\label{FigDiel-time}
	\end{figure}
	
The figure \ref{FigDiel-time} shows the calculated dielectric constant in time to 55 nanoseconds. As observed, this long simulation time is required to reach the stability shown because within the first 30 ns the value fluctuates, generating errors of 10\% or more.

\newpage


\subsubsection{ Self diffusion coefficient, D.}

The self diffusion coefficients of the cation, D+, and anion, D-, were determined using the Einstein relation, which connects the mean-square displacement (MSD) to the self diffusion in the limit of long times, through
\begin{equation}
\label{dif}
D = \lim_{t \to \infty } \frac{1}{6t}<\sum\limits_{i=1}^N[\vec { r_i}(t) - \vec{r_i} (0)]^2 >,
\end{equation}
where, in the calculations, we used the position of the center of mass of each ion, ${\vec r}_i$ \cite{Fuentes-Azcatl2014-ez}.

\begin{figure}[H]
\vspace{2.0 cm}
		\centering
	 {\includegraphics[clip,width=12cm,angle=-90]{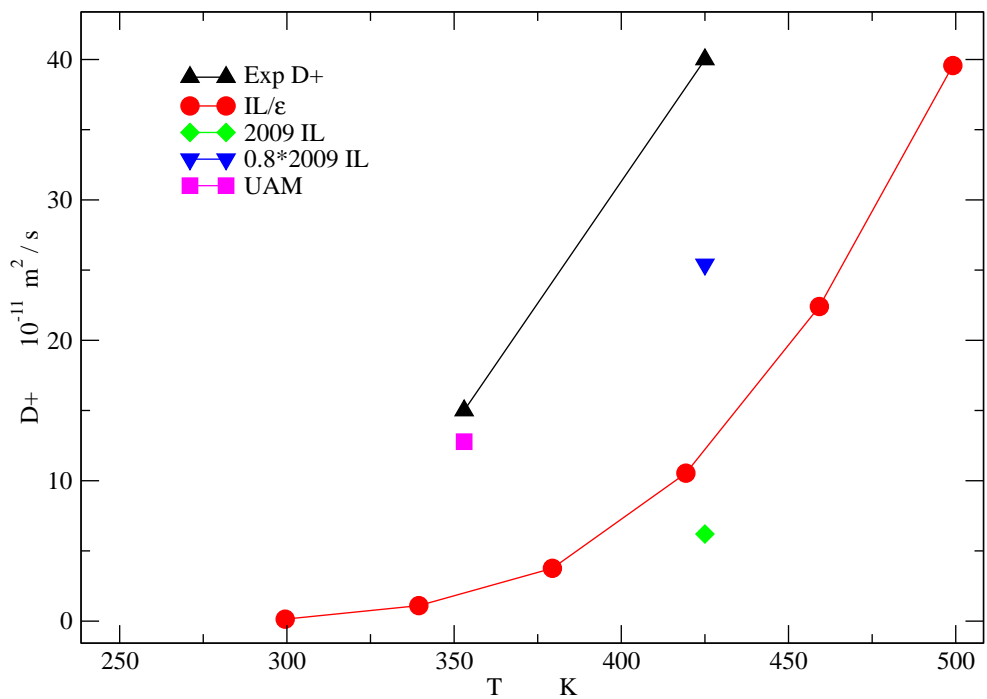}}
		
	 \caption{self diffusion coefficient of cation  in the system [bmim] [Tf2N] at different temperatures. Black triangles are the experimental data \cite{}. the red circles are the result of IL/$\epsilon$, the blue down triangle is reported by Doherty et al \cite{Doherty2017-st} , the green diamon is reported by Sambasivarao et al \cite{Sambasivarao2009-pt} and magenta square was reported by \textcolor{blue}{Mart\'inez-Jim\'enez et al. \cite{MARTINEZJIMENEZ2021115488}}}
		\label{FigDif+}
	\end{figure}

\begin{figure}[H]
\vspace{2.0 cm}
		\centering
	 {\includegraphics[clip,width=12cm,angle=-90]{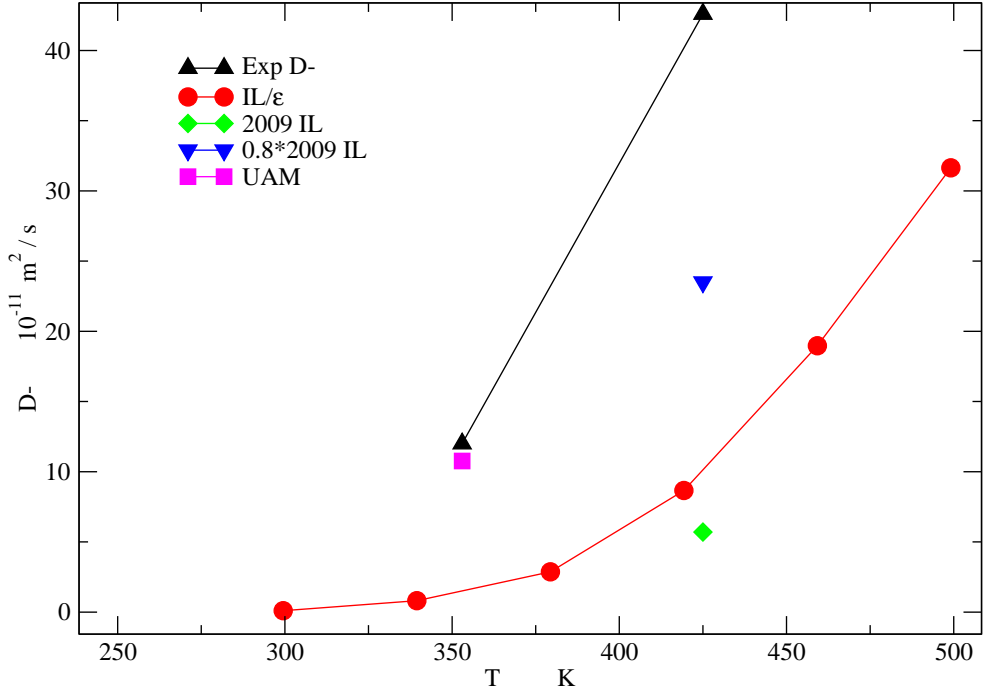}}
		
	 \caption{self diffusion coefficient of anion in the system [bmim] [Tf2N] at different temperatures. Black triangles are the experimental data \cite{}. the red circles are the result of IL/$\epsilon$, the blue down triangle is reported by Doherty et al \cite{Doherty2017-st} , the green diamon is reported by Sambasivarao et al \cite{Sambasivarao2009-pt} and magenta square was reported by Mart\'inez-Jim\'enez et al. \cite{MARTINEZJIMENEZ2021115488}}
		\label{FigDif-}
	\end{figure}

Typically, the absolute values of the ion diffusion constants are 2 orders of magnitude smaller than those of water \cite{spce}, i.e., 10$^{-11}$ m${^2}$ /s.
Figures \ref{FigDif+} and \ref{FigDif-} show the data obtained for D+ and D- with the new Il/$\epsilon$ FF, and those reported by Doherty et al \cite{Doherty2017-st}, Sambasivarao et al \cite {Sambasivarao2009-pt} and Mart\'inez-Jim\'enez et al \cite{MARTINEZJIMENEZ2021115488}. The later is the one that best reproduces the experimental value, which is not a surprise since it was an objective property in the reparameterization of the UAM FF, as well as the experimental density \cite{MARTINEZJIMENEZ2021115488}.

\subsubsection{Isothermal compressibility, k$_T$. }

The evaluation of how pressure affects density at constant temperature is determined by calculating the isothermal compressibility, k$_T$. The calculation of this property is done by means of the fluctuations in the volume during the simulation in the NPT ensemble, as shown in the equation\ref{kT}
\begin{equation}
\label{kT}
k_T= \frac{1}{\rho} \left(\frac{\partial \rho}{\partial P}\right)_T = \frac {<V^2> - <V>^2}{k_B T<V>}
\end{equation}

\begin{figure}[H]
\vspace{2.0 cm}
		\centering
	 {\includegraphics[clip,width=12cm,angle=-90]{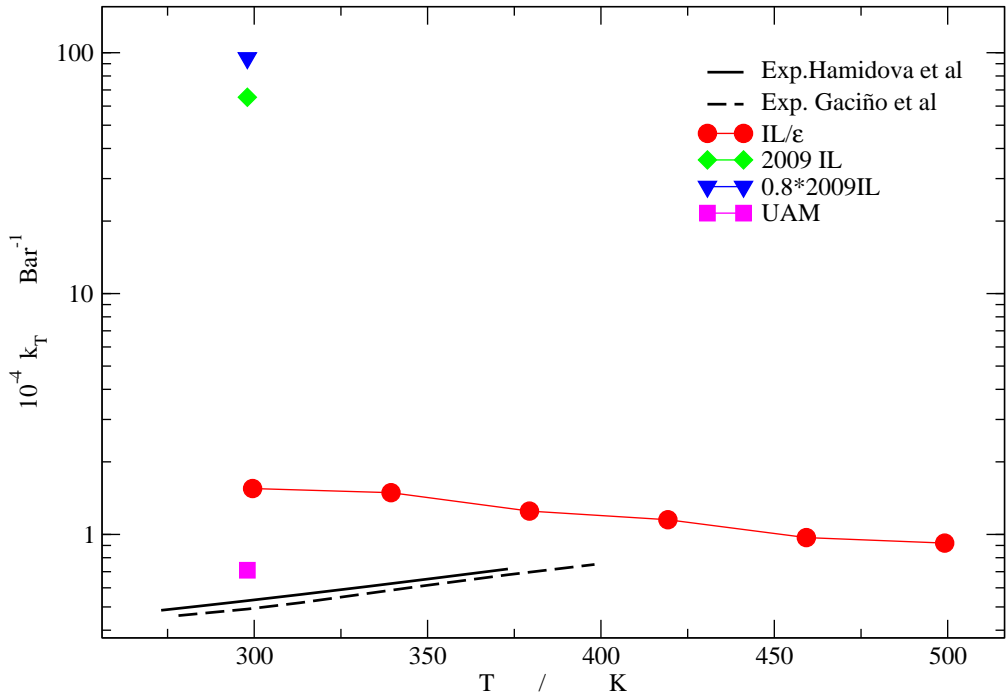}}
		
	 \caption{Isothermal compressibility of the system [Bmim] [Tf2N] at different temperatures. Black line reported by Hamidova et al \cite{hamidova} and black dash line reported by Gaciño et al\cite{GACINO2015124} are the experimental data, the red circles are the result of IL/$\epsilon$,  the blue down triangle  by Doherty et al \cite{Doherty2017-st}, the green diamon by Sambasivarao et al \cite{Sambasivarao2009-pt}, magenta square by Mart\'inez-Jim\'enez et al \cite{MARTINEZJIMENEZ2021115488} all were calculated in this work.}
		\label{FigKt}
	\end{figure}

The data obtained in this work are shown in figure \ref{FigKt}. The value produced by the 2009IL FF and that of the 0.8*2009IL FF are far from the experimental value whereas the UAM FF produces a value with an error of 31\% greater the experimental value at 300 K and 1 bar while the new IL/$\epsilon$ FF has an error of 172\% at those conditions. Several values of k$_T$ were calculated with the new FF IL/$\epsilon$ to analyze its behavior when the temperature rises. We see that it decreases and in general keeps a behavior with little change, while the experiments indicate that it must increase, 
suggesting that in this new FF the attractive interactions or hydrogen bonding are overestimated and the system do not expand as temperature rises. For most simple liquids, the compressibility decreases with decrease in temperature as the volume fluctuation decreases, whereas in our case the behavior is similar to water below about 320 K, i.e., compressibility is higher when temperature decreases.

\subsection{Mixture of IL/$\epsilon$ - H$_2$O.}

By reparameterizing the [Bmim][Nf$_2$T] system, the performance of this new FF is evaluated in a mixture with water, the FBA/$\epsilon$\cite{fbae} flexible 3-sites water model is used, this model improves the reproduction of experimental data at various thermodynamic conditions with respect to other rigid or flexible three sites models with equal interaction potentials.\cite{tip4pef}. This evaluation process has been carried out successfully in the case of NaCl\cite{nacle} and KBr\cite{kbre}, which originates a systematic test for ionic species.

As reported in the work of Fuentes et al\cite{Fuentes-Azcatl2021} with three different water force fields, the reproduction of the thermodynamic properties of the mixture is very similar in low and high concentrations regardless of the water model used, for this reason in this work, only the FAB/$\epsilon$ model was used, which performs well at the pressure and temperature conditions studied in this work.

\subsubsection{Density, $\rho$.}

\begin{figure}[H]
\vspace{2.0 cm}
		\centering
	 {\includegraphics[clip,width=12cm,angle=0]{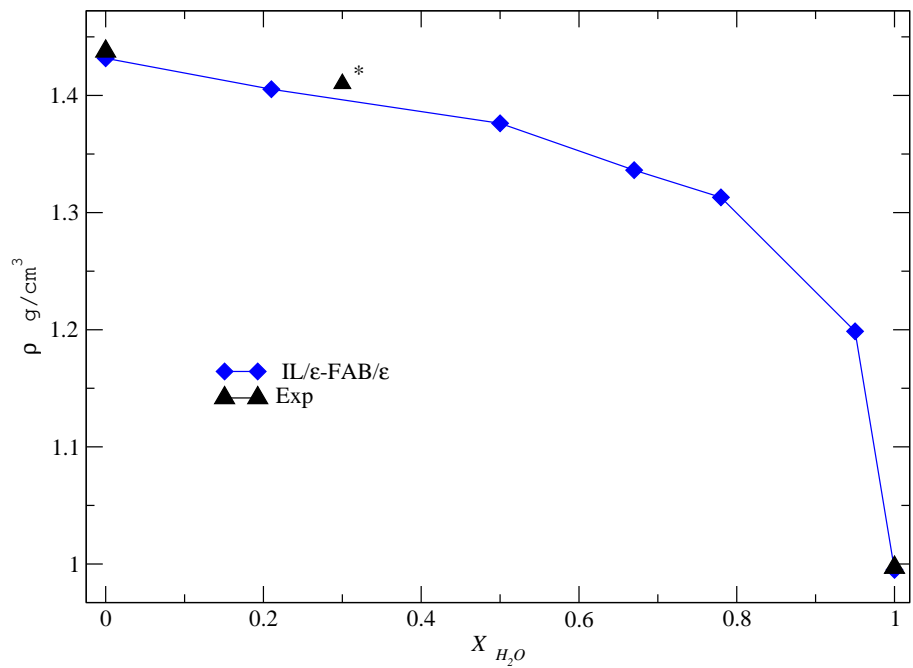}}
		\caption{Density of the water–IL solution as a function of the molar fraction of water at 1 bar of pressure. Black triangles are the experimental data \cite{MARTINS2016188}, black triangle-asterisk is the experimental data reported by Jacquemin et al. with water-saturated sample \cite{B513231B}, the blue diamonds are the result using FAB/$\epsilon$ model, all calculated in this work}
		\label{Dens-h2o}
	\end{figure}

A second reason to use the FBA/$\epsilon$ water model, is because among the three-site models, it is the one that best reproduces the experimental values of density, heat of vaporization, surface tension and dielectric constant \cite{fbae}. So, these properties were evaluated for the mixture of Il/$\epsilon$ with H$_2$O. The result of the density with respect to the temperature at a pressure of 1 bar can be seen in the figure \ref{Dens-h2o}, where the values of the pure substances reproduce the experimental data. It is observed that in the molar fraction of X$_ {H_2O}$=3, the experimental data is reproduced. Therefore, our results suggest that is highly likely that the experimental data of the mixture will be well reproduced by the new FF.

\subsubsection{Heat of Vaporization, ${\Delta}H_{vap}$.} 
\begin{figure}[H]
\vspace{2.0 cm}
		\centering
	 {\includegraphics[clip,width=12cm,angle=-90]{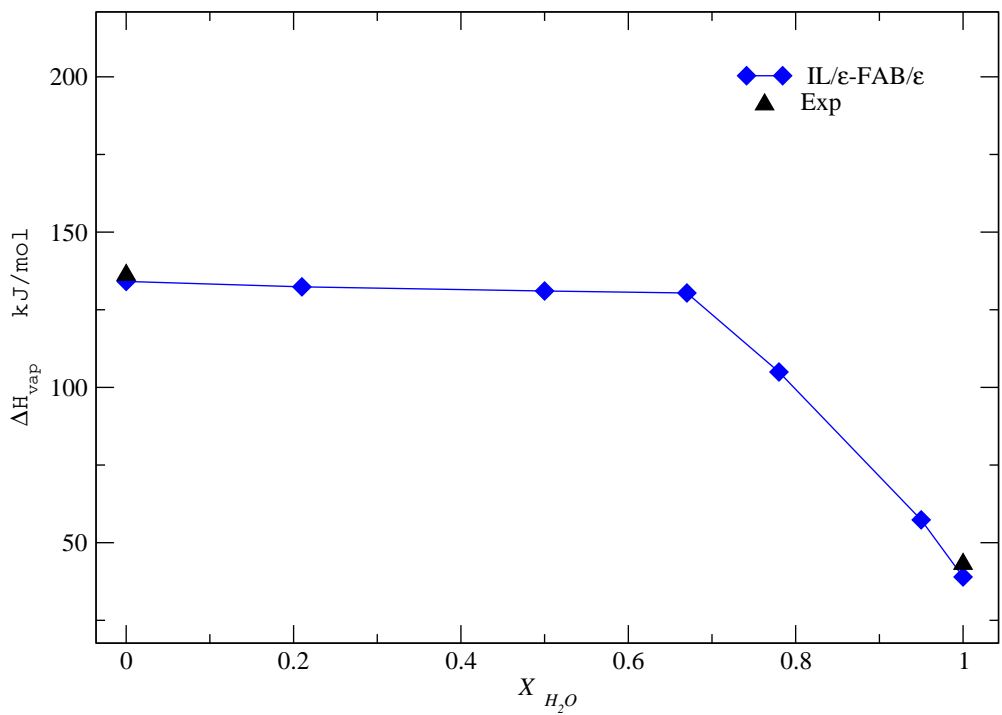}}
			 \caption{$\Delta$Hvap of the system with 
respect to the fraction of water in the solution Il-water. Both at 1 bar of pressure and 300 K of temperature. Black triangles are the experimental data \cite{Zaitsau2006-bh}, the blue diamonds are the result using FAB/$\epsilon$ model, all calculated in this work}
		\label{DH-h2o}
	\end{figure}

In the experiments carried out by Zaitsau et al\cite{Zaitsau2006-bh} they demonstrated the equilibrium of thermal stability and volatility for [Bmim][NTf$_2$ ] In fact, they made measurements of vapor pressures in a temperature interval large enough for an adequate evaluation of the thermodynamic parameters, such as the change of enthalpy from the gas phase to the liquid phase $\Delta H_{vap}$. In the figure \ref{DH-h2o} it can be seen how both models, the one for the ionic liquid Il/$\epsilon$ and that of water, the FBA/$\epsilon$, reproduce the experimental data with a minimum percentage of error. Then, from calculations at different proportions of water and Il, we calculated the values of $\Delta H_{vap}$ of the mixture, which maintain a reasonable behavior and adapt well to the range of values of the pure components.
\subsubsection{Surface tension, $\gamma$.}
\begin{figure}[H]
\vspace{2.0 cm}
		\centering
	 {\includegraphics[clip,width=12cm,angle=-90]{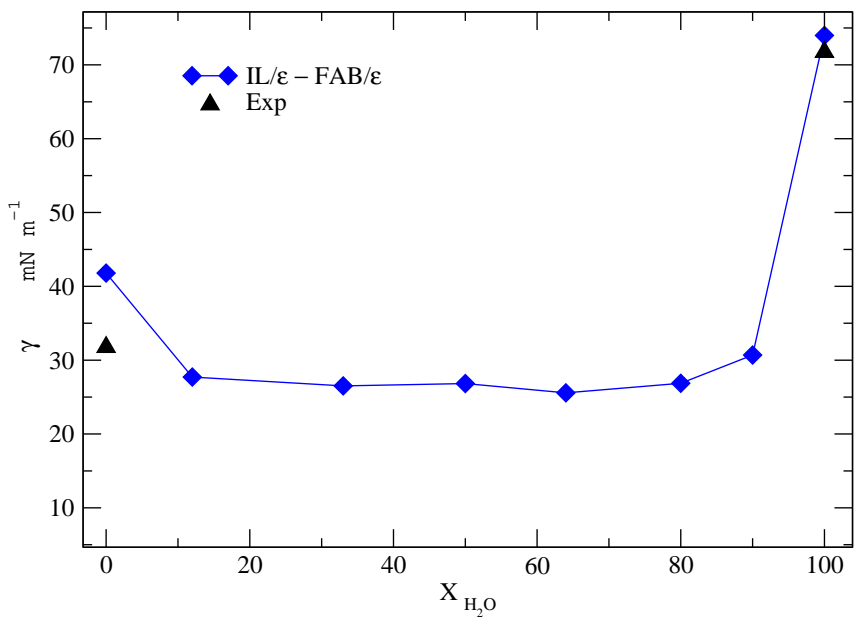}}
			 \caption{The surface tension of the system with respect to the fraction mol of water in the solution IL-water. Black triangles are the experimental data \cite{hamidova, NIST}, the blue diamonds
are the result using FAB/$\epsilon$ model with the IL/$\epsilon$, all calculated in this work}
		\label{ST-h2o}
	\end{figure}

The surface tension of the system when the fraction mol of water increases, X$_{H_2O}$, is shown in figure \ref{ST-h2o}. The new FF Il/$\epsilon$ overestimated the experimental value of the surface tension with an error of 23.8\%, but when the mol fraction of water increases, the trend is almost the same until X$_{H_2O}$=0.9. At this concentration, the surface tension increased and reach the experimental value of pure water with an error of 3\%, indicating that the ionic liquid governs the surface tension up to the limit of X$_{H_2O}$=0.9.

\subsubsection{ Dielectric constant, $\epsilon$.}

\begin{figure}[H]
\vspace{2.0 cm}
		\centering
	 {\includegraphics[clip,width=12cm,angle=-90]{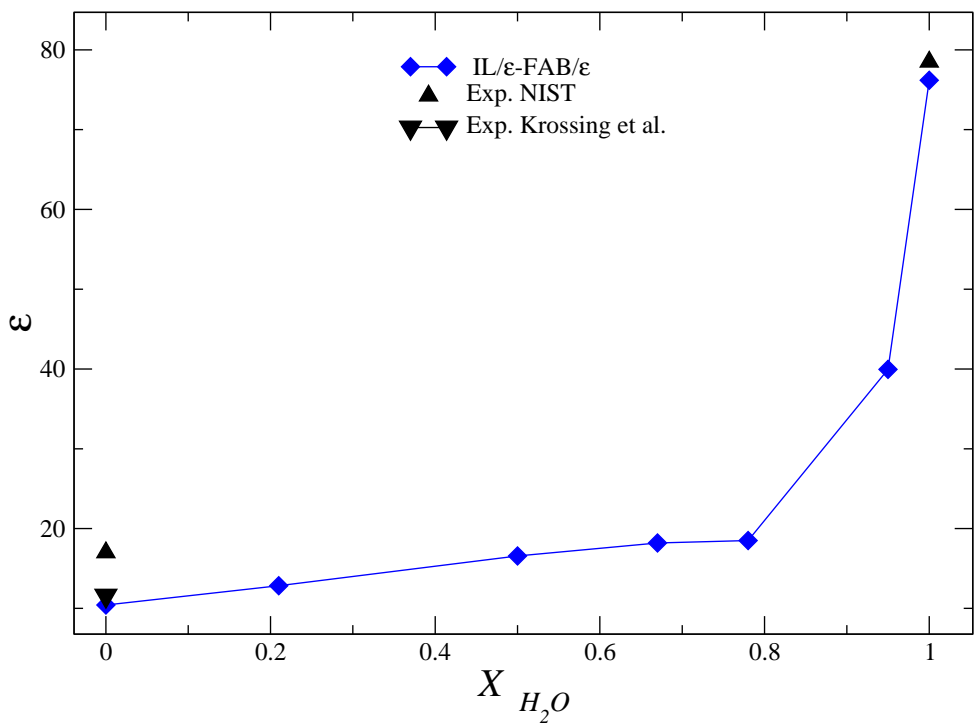}}		
	 \caption{ Dielectric constant of the water–IL solution as a
function of the molar fraction of water at 1 bar of pressure and 298 K of temperature. Black triangles
are the experimental data \cite{NIST}. The blue
diamonds are the result using
FAB/$\epsilon$ model, all calculated in this work}
		\label{Dcte-h2o}
	\end{figure}

The calculation of the dielectric constant of the system is shown in figure \ref{Dcte-h2o}. The error the FF Il/$\epsilon$ carries is 35\% respect to the NIST database \cite{NIST} and 5\% respect to that reported by Krossing et al \cite{Krossing2006-bc}. The FAB/$\epsilon$ FF of water at these conditions has a 3.8\% of error respect to the experimental data, then taking this into account, the values obtained for the mixture have that percentage of error. Even so, the electrostatic interactions generated by the Il govern the behavior of the mixture up to X$_{H_2O}$=0.8, where the greater presence of water causes the system to increase the value of $\epsilon$, to finally reach a value very close to the experimental one with the error indicated above.

\section {Conclusion}

In this work, the ionic liquid [Bmim][Nf$_2$T] was reparameterized, which improved the reproduction of density, $\Delta$H$_{vap}$, and the dielectric constant. The isothermal compressibility, self-diffusion coefficient and surface tension are also reported, which improve the reproduction of these properties with respect to the force fields compared here. There are still expectations and challenges to improve the force fields of the ionic liquids and have a better description of the experimental values at different conditions of pressure and temperature. There is also room to predict and eventually improve the values of the properties here reported when the Il's are combined  with other molecules. The reported the properties of the mixture of IL and water, calculated with the new force field at different concentrations, being certain that each model reproduce the experimental properties separately, and obtained a consistent behavior for the values of the mixture. These findings suggest that the predicted properties would be in good agreement with the experimental data. The data here reported for density, $\Delta$H$_{vap}$, surface tension and dielectric constant, leave a broader picture of what the new force fields can reproduce. With force fields capable of reproducing experimental properties with less error, there is greater certainty that they will reproduce values at different thermodynamic states and concentrations where there are still no experimental data. 

\section {Acknowledgements}
 
RFA thanks CONACYT for a postdoctoral fellowship. The authors acknowledge the computer resources and support provided by
Laboratorio Nacional de Superc\'omputo del Sureste de M\'exico,
CONACYT network of national laboratories.
Support from VIEP-BUAP through project CA-191-GOMM (2023)
is also acknowledged.

\bibliography{achemso}

\end{document}